# Thermal formation of ammonium carbamate on the surface of laboratory analogues of carbonaceous grains in protostellar envelopes and planet-forming disks

Alexey Potapov[1], Cornelia Jäger[1], and Thomas Henning[2]

[1]*Laboratory Astrophysics Group of the Max Planck Institute for Astronomy at the Friedrich Schiller University Jena, Institute of Solid State Physics, Helmholtzweg 3, 07743 Jena, Germany, email: alexey.potapov@uni-jena.de*
[2]*Max Planck Institute for Astronomy, Königstuhl 17, D-69117 Heidelberg, Germany*

**Abstract**

The catalytic role of dust grain surfaces in the thermal reaction $CO_2 + 2NH_3 \rightarrow NH_4^+NH_2COO^-$ was recently demonstrated by our group. The rate coefficients for the reaction at 80 K on the surface of nanometre-sized carbon and silicate grains were measured to be up to three times higher compared to the reaction rate coefficients measured on KBr. In this study, the reaction was performed on carbon grains and on KBr in the extended temperature range of 50 - 80 K and with the addition of water ice. The reaction activation energy was found to be about 3 times lower on grains compared to the corresponding ice layer on KBr. Thus, the catalytic role of the dust grain surface in the studied reaction can be related to a reduction of the reaction barrier. Addition of water to $NH_3:CO_2$ ice on grains slowed the reaction down. At the $H_2O:CO_2$ ratio of 5:1, the reaction was not detected on the experimental timescale. This result calls into question the thermal formation of ammonium carbamate in dense molecular clouds and outer regions of protostellar and protoplanetary environments with dominating water ice mantle chemistry. However, it can still happen in inner regions of protostellar and protoplanetary environments in crystalline ices.

## 1. Introduction

Formation pathways of complex organic molecules (COMs) in interstellar and circumstellar media are of great interest to understand our astrochemical heritage (Caselli & Ceccarelli 2012). Chemical processes leading to the formation of molecules in space can be divided into gas phase and grain surface reactions. The later occur on carbonaceous and siliceous dust grains and molecular ices that provide a surface where reactants can meet and dissipate excess energy to a third body. Grains can also provide molecular building blocks, e.g. functional groups and atoms, and serve as catalyst in chemical reactions. The main triggers of grain surface chemistry are UV irradiation, cosmic rays, thermal processing, atom addition, and reactions involving radicals.

It is generally accepted that dust grains in cold and dense cosmic environments starting from molecular clouds and developing into protostellar envelopes and protoplanetary disks are covered (in protostellar and protoplanetary environments beyond their snow lines) by molecular ices. The ice mantel is typically considered to be thick (hundreds of monolayers). In this case, chemical reactions occurring in such ices are barely sensitive to the surface of dust and bare dust grains are considered to be important only in diffuse and translucent molecular clouds and in circumstellar disks beyond the sublimation radius. However, recent laboratory studies showed that, due to the high porosity of dust grain analogues, their ice coverage can be much thinner than previously assumed (Potapov, Jäger, & Henning 2018) leading to chemical reactions not in thick ices but in a monolayer or a few monolayers of ice on the surface of dust (Potapov et al. 2019b).

In the laboratory, reactions on the surface of reliable cosmic dust grain analogs are studied rarely. Except for the formation of $H_2$ and $H_2O$ on silicate and carbon surfaces (Pirronello et al. 1997; Pirronello et al. 1999; Perets et al. 2007; Jing et al. 2011; Gavilan et al. 2014; He & Vidali 2014; Wakelam et al. 2017), the only study demonstrating a pathway to COMs was devoted to the formation of $H_2CO$ and $CH_3OH$ on amorphous carbon grains (Potapov et al. 2017). The catalytic role of the surface of amorphous silicate and carbon grains in the formation of COMs using the $CO_2 + 2NH_3 \rightarrow NH_4^+NH_2COO^-$ thermal reaction was recently experimentally demonstrated (Potapov, et al. 2019b). In the past, this reaction was studied in ices on standard laboratory substrates only (Hisatsune 1984; Bossa et al. 2008; Noble et al. 2014; Rodriguez-Lazcano et al. 2014; Ghesquiere et al. 2018).

Ammonium carbamate $NH_4^+NH_2COO^-$ is a complex organic molecule, which can initiate a network of prebiotic reactions. Such reactions are expected to take place in interstellar and circumstellar environments (Brack 1999; Caselli & Ceccarelli 2012) leading to the formation of biomolecules required for the origin of life on Earth (Oro 1961; Pearce et al. 2017). The aim of the present study was to investigate the formation of ammonium carbamate on the surface of dust grains at different temperatures corresponding to protostellar envelopes and planet-forming disks but below the desorption temperatures of the reactants and with an addition of water ice. We also aimed at clarifying the catalytic role of grains not revealed in our previous study (Potapov, et al. 2019b).

## 2. Experimental part

The experiments were performed using the INterStellar Ice Dust Experiment (INSIDE) setup presented in details elsewhere (Potapov, Jäger, & Henning 2019a). The base pressure in the main chamber of the setup is a few $10^{-11}$ mbar allowing to reproduce physical conditions in dense cosmic environments, such as molecular clouds, protostellar envelopes, and planet-forming disks and to perform clean experiments without additional adsorption of species (mainly water) from the chamber volume. Nanometre-sized amorphous carbon grains were produced in a laser ablation setup (Jäger et al. 2008; Potapov, et al. 2018) by pulsed laser ablation of a $^{13}C$ solid target and subsequent condensation of the evaporated species in a quenching atmosphere of 4 mbar He and $H_2$. Condensed grains were deposited onto a KBr substrate. The thickness of the grain layer with the density of 1.7 g cm$^{-3}$ was 40 nm controlled by a microbalance.

After the deposition of grains, the sample was extracted from the deposition chamber, fixed on a sample holder, and inserted into the pre-chamber of INSIDE, where it was annealed at 200º C for two hours to remove possible surface contamination by molecular adsorbates. After annealing, the sample holder was moved to the ultra-high vacuum chamber of INSIDE and fixed on the coldhead. The pure KBr substrate used in this study was also annealed. $NH_3$ and $CO_2$ (purity 5.5) with a ratio of 4:1 were deposited on the surfaces of KBr or carbon grains at 15 K. The gases were introduced through two separated gas lines. The deposition angle was 90º. The $CO_2$ and $NH_3$ number of molecules deposited were calculated from their vibrational bands at 2342 and 1073 cm$^{-1}$ using the band strengths of $7.6\times10^{-17}$ cm molecule$^{-1}$ (Gerakines et al. 1995) and $1.7\times10^{-17}$ cm molecule$^{-1}$ (d'Hendecourt & Allamandola 1986). Additional experiments were performed with ice mixtures containing water ice. $H_2O$ was premixed with $CO_2$ in the ratios of $H_2O:CO_2$ (2:1) and (5:1). The number of $H_2O$ molecules deposited was calculated from the vibrational band around 3300 cm$^{-1}$ using the band strengths of $2\times10^{-16}$ cm molecule$^{-1}$ (Hudgins et al. 1993) and subtracting the overlapping $NH_3$ band. The deposition time in all experiments was 15 minutes meaning the constant deposition rates for $CO_2$ of about $3.4\times10^{15}$ molecule minute$^{-1}$ and $NH_3$ of about $13.6\times10^{15}$ molecule minute$^{-1}$.

After the deposition of ices, the samples were heated up to a definite temperature with the highest heating rate available, 100 K min$^{-1}$, to minimize the reaction time at lower temperatures. After heating, the initial time was set to zero, and the isothermal kinetics experiments were performed during 4 hours.

The temperature range probed was 50 - 80 K. The lower limit is due to the fact that at 50 K, the studied reaction was not observed on the experimental timescale. The higher limit is due to the noticeable thermal desorption of $CO_2$ above 80 K. The formation of the $NH_4^+NH_2COO^-$ product was monitored by the growth of its characteristic vibrational bands at 1550 and 1385 cm$^{-1}$, corresponding to the assymetric and symmetric stretching vibrations of the $COO^-$ unit, as a function of time. The column density of $NH_4^+NH_2COO^-$ was calculated from the more intense 1550 cm$^{-1}$ band using the band strength of $2\times10^{-17}$ cm molecule$^{-1}$ (Bossa, et al. 2008). IR spectra were measured using an FTIR spectrometer (Vertex 80v, Bruker) in the transmission mode with a resolution of 1 cm$^{-1}$.

The very large surface of grains, which can be two orders of magnitude larger than the nominal area on a KBr substrate (Potapov, et al. 2018) means that the number of monolayers is different on the grains and on the KBr surface. In our previous study of the $CO_2 + 2NH_3 \rightarrow NH_4^+NH_2COO^-$ reaction on grains (Potapov, et al. 2019b), it was determined that the multilayer-monolayer transition occurred at a nominal ice thickness of about 200 nm on grains while on KBr surface ice remained in the multilayer regime in the whole range of thicknesses studied, as low as 35 nm. The thickness in nm is calculated from the number of deposited molecules assuming a monolayer column density of $1\times10^{15}$ molecules cm$^{-2}$ and a monolayer thickness of 0.3 nm (Opitz et al. 2007).

In the present study, the nominal thickness of $CO_2:NH_3$ ice was fixed at 85 nm. As far as the ice thickness on dust is not known, we cannot compare to any definite thickness on KBr. Instead, we compare two systems, KBr and dust grains, covered by the same amount of ice molecules. This is much more important because we compare a typical "model" system of pure ices to the "real" system of ice on dust and show differences.

Addition of water increases the total ice thickness to 120 nm for the $H_2O:CO_2:NH_3$ (2:1:4) mixture and to 170 nm for the $H_2O:CO_2:NH_3$ (5:1:4) mixture, thus, in all experiments the monolayer regime on grains was ensured.

## 3. Results

In Figure 1, the IR spectra of a $H_2O:CO_2:NH_3$ (2:1:4) ice mixture taken directly after the deposition at 15 K and after 4 hours of the isothermal kinetics experiment at 75 K are shown. A detailed assignment of the IR absorption bands of $NH_3:CO_2$ ices is done elsewhere (Bossa, et al. 2008; Noble, et al. 2014). Two bands related to $NH_4^+NH_2COO^-$ at 1550 and 1385 cm$^{-1}$ are observable in the 75 K spectrum.

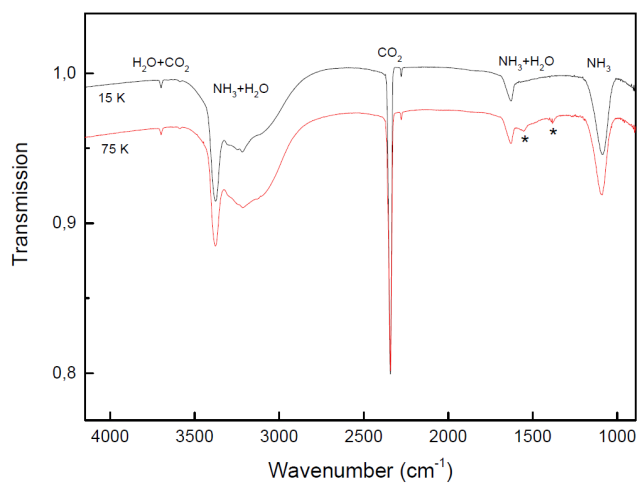

Figure 1. IR spectra taken after the deposition of a $H_2O:CO_2:NH_3$ (2:1:4) mixture on carbon grains at 15 K and after 4 hours at 75 K. Two bands related to $NH_4^+NH_2COO^-$ are marked by asterisks. The 75 K spectrum is vertically shifted for clarity.

The results of the isothermal kinetics experiments are shown in Figure 2.

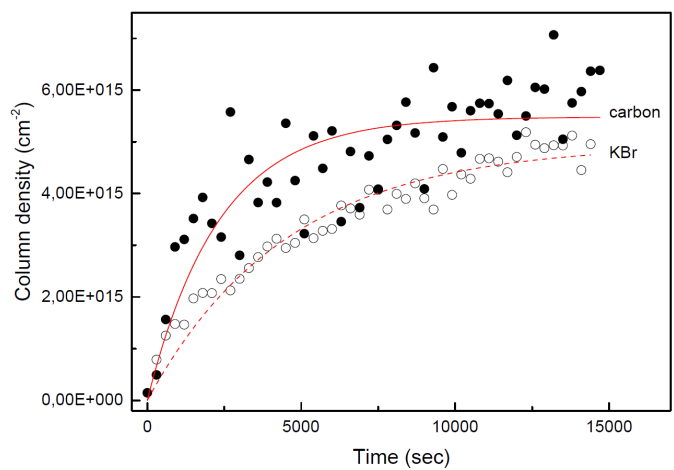

Figure 2. Derived time dependences of the $NH_4^+NH_2COO^-$ column density calculated from the 1550 cm$^{-1}$ band using equation (1) for the isothermal kinetic experiments performed with $CO_2:NH_3$ 1:4 ices on carbon grains (filled circles) and on KBr (empty circles) at 75 K.

The isothermal kinetic curves for ices on grains reach a plateau as it is seen in Figure 2. One experiment on KBr was repeated on the time-scale of 6 hours and gave a similar value for the reaction rate coefficient compared to the 4-hours experiment. The growth of the $NH_4^+NH_2COO^-$ column density exhibits pseudo- first-order reaction kinetics and the reaction rate coefficients were determined solving the kinetic equation (Noble, et al. 2014):

$$(NH_4^+NH_2COO^-)(t) = (CO_2)_0 \times (1 - e^{-kt}) \quad (1)$$

In Figure 3 we present the temperature dependencies of the reaction rate coefficient for three systems studied, i.e. $CO_2:NH_3$ on carbon grains, $H_2O:CO_2:NH_3$ on carbon grains, and $CO_2:NH_3$ on KBr. The reaction rate coefficients obtained using equation (1) are given in Table 1. In our previous study (Potapov, et al. 2019b), we showed that surface catalysis on silicate and carbon grains accelerates the kinetics of the $CO_2 + 2NH_3 \rightarrow NH_4^+NH_2COO^-$ reaction at a temperature of 80 K compared to the corresponding ice layer on KBr. The results of the present study confirm this finding and extend it to lower temperatures. A factor of two between the reaction rate coefficients on grains and on KBr obtained for the given ice thickness stays constant in the temperature range 65 - 80 K. At 60 K, the reaction is observable on grains in contrast to the ice layer on KBr. Ammonium carbamate was not detected in the $CO_2:NH_3$ ices on carbon grains and on KBr at 50 and 60 K respectively, and in the water-containing ice at 60 and 65 K.

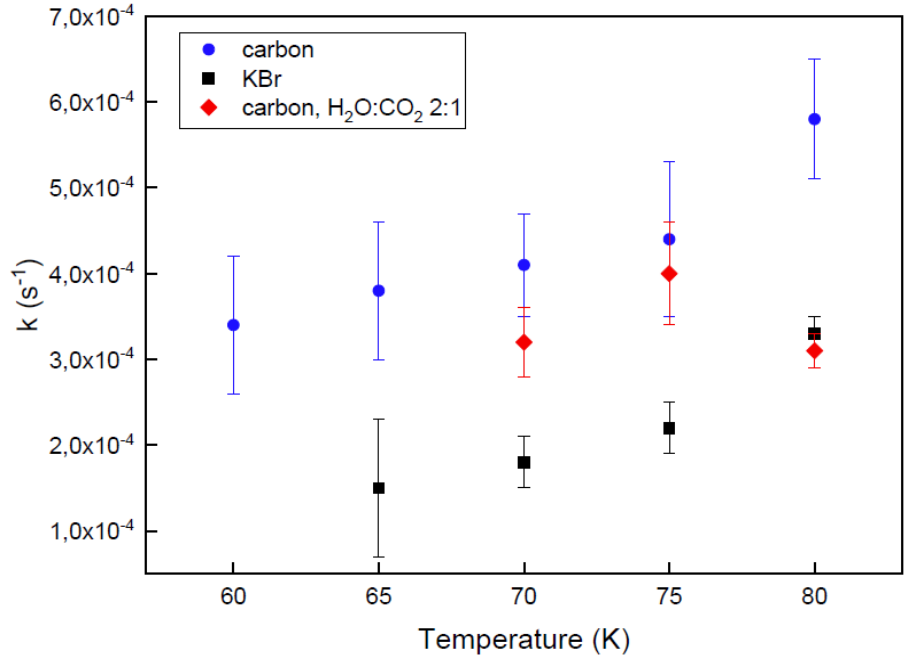

Figure 3. Temperature dependencies of the reaction rate coefficient: for $CO_2:NH_3$ (1:4) on carbon grains (circles) and on KBr (squares), and for $H_2O:CO_2:NH_3$ (2:1:4) on carbon grains (diamonds).

Table 1. Reaction rate coefficients (in $10^{-4}$ s$^{-1}$) for different temperatures and samples.

| T (K) | KBr CO$_2$:NH$_3$ (1:4) | Carbon grains CO$_2$:NH$_3$ (1:4) | Carbon grains H$_2$O:CO$_2$:NH$_3$ (2:1:4) | Carbon grains H$_2$O:CO$_2$:NH$_3$ (5:1:4) |
|---|---|---|---|---|
| 50 | | no reaction* | | |
| 60 | no reaction* | 3.4 ± 0.8 | no reaction* | |
| 65 | 1.4 ± 0.9 | 3.8 ± 0.9 | no reaction* | |
| 70 | 1.8 ± 0.3 | 4.1 ± 0.6 | 3.2 ± 0.4*** | no reaction* |
| 75 | 2.2 ± 0.3 | 4.4 ± 0.9 | 4.0 ± 0.6*** | no reaction* |
| 80 | 3.3 ± 0.2** | 5.8 ± 0.7** | 3.1 ± 0.3*** | |

*ammonium carbamate was not detected on the experimental timescale

**(Potapov, et al. 2019b)

***the H$_2$O:CO$_2$:NH$_3$ (2:1:4) mixtures were measured twice because of the unexpected behavior of the reaction rate dependence. The reaction rate coefficients obtained in two experiments differ by about 10%. The values in the table are average values.

The addition of water slows down the reaction. This effect was observed in the previous studies (Noble, et al. 2014; Rodriguez-Lazcano, et al. 2014), however, it is hard to make a quantitative or even a qualitative comparison because the influence of water can be very different for multilayer (water ice matrix) and monolayer regimes. In the present study, for the H$_2$O:CO$_2$:NH$_3$ (2:1:4) system, intermediate reaction rate coefficients between those obtained for grains and for KBr at 70 and 75 K were observed. The coefficient at 80 K is lower than the one at 75 K. The decrease of the reaction rate coefficient may be linked to the transformation of the water ice structure from high-density to low-density ice finalising around 80 K (Narten, Venkatesht, & Rice 1976; Jenniskens et al. 1995) or to the structural changes in ammonia ice that forms a more ordered hydrogen bonded structure around 80 K (Dawes et al. 2007). With the increased amount of water ice in the ice mixture, H$_2$O:CO$_2$:NH$_3$ (5:1:4), no reaction was observed on grains at 70 and 75 K on the experimental timescale pointing that at a high amount of water ice the reactants are isolated in water ice or strongly bound to water ice molecules and cannot meet each other on the experimental timescale.

Fitting the temperature dependencies of the reaction rate coefficients with the Arrhenius law gives experimental activation energies of 310 ± 43 K (2.6 ± 0.4 kJ mol$^{-1}$) for the reaction on KBr, 116 ± 26 K (1.0 ± 0.2 kJ mol$^{-1}$) for the reaction on carbon grains, and 234 K (2.0 kJ mol$^{-1}$) for the reaction on carbon grains in the presence of water obtained from two points, 70 and 75 K. Thus, the reaction energy barrier for the CO$_2$ + 2NH$_3$ reaction is about three times lower on grains compared to the KBr surface with an intermediate case of the $H_2O:CO_2:NH_3$ (2:1:4) system. The pre-exponential factors for all systems are low, 0.01 s$^{-1}$ for the reaction on KBr and 0.003 s$^{-1}$ for the reaction on carbon grains. In (Noble, et al. 2014), the small value of the pre-exponential factor, 0.09 s$^{-1}$, was referred to either a sequence of elementary processes presented by the reaction or a contribution from quantum tunnelling.

We have to stress that we cannot directly distinguish between reaction and diffusion processes from our measurements. However, we can try to compare the timescales meaning that we measure the slower process. The experimental diffusion rates $D$ for $CO_2$ and $NH_3$ along pores of amorphous water ice were measured to be about 10$^{-14}$ cm$^2$ s$^{-1}$ or higher (Mispelaer et al. 2013; Ghesquiere et al. 2015). The diffusion rate $k_{diff}$ was calculated using the expression $D = k_{diff} \times d^2/2$, where $d$ was the thickness of water ice (Mispelaer, et al. 2013). In our case of a sub-monolayer coverage on grains, for $d$ we can use 0.3 nm, the distance between two nearest adsorption sites (He, Emtiaz, & Vidali 2018), multiplied by $N$, the number of hops. $N$ is not known because of the unknown surface area of grains. However, taking even 100 hops, which is not really reasonable (typically a few hops are needed), the diffusion time is about 2 seconds, much shorter than our experimental timescale of $1/k$ giving the time of a few thousand seconds.

Thus, we can conclude that the measured process is much slower than the diffusion process and should be the reaction process. Consequently, the measured activation energies are equivalent to reaction energies of the thermal reaction between $NH_3$ and $CO_2$. We have to admit that the main problem of our comparison is that the diffusion was measured along pores of amorphous water ice and can be quite different on the surface of grains. However, there is no reason to suppose that the diffusion on grains can be orders of magnitude slower.

As it was stated in our previous study (Potapov, et al. 2019b), the increased reaction rates for grains compared to those measured for the $NH_3:CO_2$ ices on KBr can be explained by an increased diffusion rate of species, by the formation of an intermediate weakly bound $CO_2$-$NH_3$ complex that helps to overcome the reaction barrier or by a reduction of the reaction barrier. The new results allow us to conclude that the main catalytic effect of the grain surface is the reduction of the reaction barrier. In the IR spectra, we do not see any evidence for the formation of an intermediate complex. Such a complex was detected in $CO_2:NH_3$ (1:1) ices (Bossa, et al. 2008) but not in the $CO_2:NH_3$ ices with an excess of $NH_3$ (Noble, et al. 2014). Probably, more $CO_2$ relative to $NH_3$ is needed for its formation. The diffusion may play a role, when structural transformations of the water ice matrix take place. Such result was obtained previously for $H_2O:CO_2:NH_3$ ices at temperatures corresponding to the transformation of water ice structure from amorphous to crystalline state (Ghesquiere, et al. 2018).

## 4. Astrophysical Discussion

Our results are important for a general understanding of the chemical processes on the surface of interstellar and circumstellar dust grains. The catalytic effect of grain surfaces is clearly demonstrated by our studies of the $CO_2 + 2NH_3 \rightarrow NH_4^+NH_2COO^-$ reaction. It is possible that this effect could lead to an enhanced solid-state reactivity for reactions involving other neutral molecules or radicals (one of the main triggers of grain surface chemistry) and has to be taken into account in future astrochemical models. Laboratory experiments on astrochemically relevant reactions on cosmic grain analogues are necessary to reveal the role and to quantify the catalytic effect of grain surfaces in these processes.

It is well known from astronomical observations that water is the main component of ices in molecular clouds and colder regions of protostellar envelopes and protoplanetary disks. The abundances of $CO_2$ and $NH_3$ molecules in these ices vary from a few percent to a few tens of percent relative to $H_2O$ (Gibb et al. 2004; Boogert et al. 2011). Thus, the $H_2O:CO_2:NH_3$ (5:1:4) ice is probably the most reliable (from the ices studied in this work) analogue of cosmic ices in these environments. The reaction was not detected for such ice mixtures on grains on the experimental timescale in contrast to the other systems, i.e. $H_2O:CO_2:NH_3$ (2:1:4) and $CO_2:NH_3$ (1:4), which are less relevant. This result calls into question the thermal formation of ammonium carbamate in interstellar and circumstellar ices, at least at temperatures up to 80 K. The structural transformation of the water ice matrix from amorphous to crystalline taking place at higher temperatures (140 K on the laboratory timescale) can change the situation dramatically. Much higher reaction rate coefficients for the $CO_2 + 2NH_3$ reaction in the water ice matrix at 120 - 140 K (Ghesquiere, et al. 2018) compared to the $CO_2:NH_3$ ice at 70 - 90 K (Noble, et al. 2014) have been obtained. It is worth to note that ammonium carbamate can be also formed in $CO_2:NH_3$ ices at low temperatures by UV irradiation, however, the main product in this case is carbamic acid $NH_2COOH$. In the thermal reaction, the ammonium carbamate:carbamic acid ratio is 1:1, while the UV irradiation at 10 K produces a ratio of 1:28 (Bossa, et al. 2008). However, UV irradiation at higher temperatures may lead to a higher formation rate of ammonium carbamate - a goal for our next set of experiments.

For astronomical observations, solid ammonium carbamate should probably be searched in circumstellar environments at temperatures higher than 100 K using the IR bands at 1550 and 1385 $cm^{-1}$. In the gas phase, ammonium carbamate or its daughter molecule, carbamic acid (what is more probable according to the mass spectra (Noble, et al. 2014; Potapov, et al. 2019b)), may be detected above its sublimation temperature of 240 K. The problem here is the lack of high-resolution spectra of ammonium carbamate and carbamic acid due to their instability in the gas phase at room temperature. One possibility to obtain high-resolution spectroscopic signatures of such species is to synthesize them in ices and then detect them above the surface of ices immediately after their desorption. Recently, the first proof-of-principle studies of this new experimental approach were presented (Yocum et al. 2019; Theulé et al. 2020). This novel approach will allow a detection of new COMs in the interstellar and circumstellar environments leading to a deeper understanding of astrochemical networks.

5. **Summary**

The formation of ammonium carbamate by the thermal reaction $CO_2 + 2NH_3$ on the surface of circumstellar carbonaceous grain analogues was studied. The catalytic effect of the carbon grain surface on the reaction rate is clearly observed in the temperature range of 60 - 80 K. The reaction activation energy on grains is about three times lower compared to the ice layer on KBr. Thus, the catalytic role of the surface can be related to a reduction of the reaction barrier. Addition of water ice in the ratio of $H_2O:CO_2$ (2:1) slows the reaction down. Addition of water ice in the ratio of $H_2O:CO_2$ (5:1), more appropriate to cold astrophysical environments, leads to non-detection of the reaction product on the experimental timescale. An astronomical search for ammonium carbamate in the solid state is probably more promising at temperatures above 100 K, where water ice undergoes the structural transformation from amorphous to crystalline.

**Acknowledgments**

We thank Patrice Theulé for fruitful discussions. We would like to thank also two anonymous reviewers for questions, suggestions, and corrections that helped to improve the manuscript. This work was supported by the Research Unit FOR 2285 "Debris Disks in Planetary Systems" of the Deutsche Forschungsgemeinschaft (grant JA 2107/3-2). TH acknowledges support from the European Research Council under the Horizon 2020 Framework Program via the ERC Advanced Grant Origins 83 24 28.

**References**

Boogert, A. C. A., et al. 2011, Astrophys J, 729
Bossa, J. B., Theule, P., Duvernay, F., Borget, F., & Chiavassa, T. 2008, Astron Astrophys, 492, 719
Brack, A. 1999, Adv Space Res-Series, 24, 417
Caselli, P., & Ceccarelli, C. 2012, Astronomy and Astrophysics Review, 20, 56
d'Hendecourt, L. B., & Allamandola, L. J. 1986, Astron Astrophys Sup, 64, 453
Dawes, A., et al. 2007, J Chem Phys, 126
Gavilan, L., Lemaire, J. L., Vidali, G., Sabri, T., & Jaeger, C. 2014, Astrophys J, 781, 79
Gerakines, P. A., Schutte, W. A., Greenberg, J. M., & van Dishoeck, E. F. 1995, Astron Astrophys, 296, 810

Ghesquiere, P., Ivlev, A., Noble, J. A., & Theule, P. 2018, Astron Astrophys, 614, A107
Ghesquiere, P., Mineva, T., Talbi, D., Theule, P., Noble, J. A., & Chiavassa, T. 2015, Phys Chem Chem Phys, 17, 11455
Gibb, E. L., Whittet, D. C. B., Boogert, A. C. A., & Tielens, A. G. G. M. 2004, Astrophys J Suppl S, 151, 35
He, J., Emtiaz, S. M., & Vidali, G. 2018, Astrophys J, 863, 156
He, J., & Vidali, G. 2014, Astrophys J, 788, 50
Hisatsune, I. C. 1984, Can J Chem, 62, 945
Hudgins, D. M., Sandford, S. A., Allamandola, L. J., & Tielens, A. G. G. M. 1993, Astrophys J Suppl S, 86, 713
Jäger, C., Mutschke, H., Henning, T., & Huisken, F. 2008, Astrophys J, 689, 249
Jenniskens, P., Blake, D. F., Wilson, M. A., & Pohorille, A. 1995, Astrophys J, 455, 389
Jing, D. P., He, J., Brucato, J., De Sio, A., Tozzetti, L., & Vidali, G. 2011, Astrophys J Lett, 741
Mispelaer, F., et al. 2013, Astron Astrophys, 555, A13
Narten, A. H., Venkatesht, C. G., & Rice, S. A. 1976, J Chem Phys, 64, 1106
Noble, J. A., et al. 2014, Phys Chem Chem Phys, 16, 23604
Opitz, A., Scherge, M., Ahmed, S. I. U., & Schaefer, J. A. 2007, J Appl Phys, 101
Oro, J. 1961, Nature, 190, 389
Pearce, B. K. D., Pudritz, R. E., Semenov, D. A., & Henning, T. K. 2017, Proceedings of the National Academy of Sciences of the United States of America, 114, 11327
Perets, H. B., et al. 2007, Astrophys J, 661, L163
Pirronello, V., Biham, O., Liu, C., Shen, L. O., & Vidali, G. 1997, Astrophys J, 483, L131
Pirronello, V., Liu, C., Roser, J. E., & Vidali, G. 1999, Astron Astrophys, 344, 681
Potapov, A., Jäger, C., & Henning, T. 2018, Astrophys J, 865, 58
---. 2019a, Astrophys J, 880, 12
Potapov, A., Jäger, C., Henning, T., Jonusas, M., & Krim, L. 2017, Astrophys J, 846, 131
Potapov, A., Theule, P., Jäger, C., & Henning, T. 2019b, ApJL, 878, L20
Rodriguez-Lazcano, Y., Mate, B., Herrero, V. J., Escribano, R., & Galvez, O. 2014, Phys Chem Chem Phys, 16, 3371
Theulé, P., Endres, C., Hermanns, M., Zingsheim, O., Bossa, J. B., & Potapov, A. 2020, ACS Earth and Space Chemistry, 4, 86
Wakelam, V., et al. 2017, Mol Astrophys, 9, 1
Yocum, K. M., Smith, H. H., Todd, E. W., Mora, L., Gerakines, P. A., Milam, S. N., & Weaver, S. L. W. 2019, J Phys Chem A, 123, 8702